\documentclass[aip,apl,reprint,preprintnumbers]{revtex4-1}
\usepackage{graphicx}
\usepackage{dcolumn}
\usepackage{bm}
\usepackage{color}
\usepackage{sidecap}
\usepackage{amssymb}

\begin{document}


\title{Efficient photon number detection with silicon avalanche photodiodes}

\author{O. Thomas}
\author{Z. L. Yuan}
\email{zhiliang.yuan@crl.toshiba.co.uk}
\author {J. F. Dynes}
\author {A. W. Sharpe}
\author {A. J. Shields}

\affiliation{Toshiba Research Europe Ltd, Cambridge Research
Laboratory, 208 Cambridge Science Park, Milton Road, Cambridge, CB4~0GZ, UK }

\date{\today}

\begin{abstract}
We demonstrate an efficient photon number detector for visible wavelengths using a silicon avalanche photodiode. Under sub-nanosecond gating, the device is able to resolve up to four photons in an incident optical pulse. The detection efficiency at 600~nm is measured to be 73.8\%, corresponding to an avalanche probability of 91.1\% of the absorbed photons, with a dark count probability below $1.1 \times 10^{-6}$ per gate. With this performance and operation close to room temperature, fast-gated silicon avalanche photodiodes are ideal for optical quantum information processing that requires single-shot photon number detection.
\end{abstract}


\maketitle

Many applications require low-level light detection, such as biomedical imaging, time-of-flight ranging, astronomy and scientific research. In particular, highly efficient and low noise single photon detectors are a prerequisite for quantum information processing based on photonic qubits.\cite{buller10} For some advanced functions, these detectors must be able to resolve the number of photons in an optical pulse.\cite{kok07,simon07,briegel98}

For over a decade, silicon avalanche photodiodes\cite{dautet93,cova04} (Si-APD) have been the detector of choice for single photon detection at visible wavelengths due to their practicality, high quantum efficiency (QE) and low dark count noise. Their spectral response also makes them compatible with a number of quantum light sources.\cite{shields07}
However, Si-APDs are believed to be threshold devices that detect only the presence or absence of photons, but not the photon number. Multiplexing in temporal\cite{achilles03} or spatial domain\cite{godik04,yamamoto06,akiba09} can be used to approximate a photon number detector (PND), but these devices suffer from cross-talk, high dark count rates as well as reduced efficiency due to the geometrical fill-factor. Alternative technologies, such as visible light photon counters\cite{petroff87,kim99} or transition edge sensors,\cite{lita08} are capable of efficient PND, but the requirement of cryogenic cooling limits their prospect for practical use.

For single photon detection, APDs are usually biased with a dc voltage above their breakdown, such that a single photo-excited carrier can stimulate a macroscopic current by means of avalanche multiplication.  In this Geiger-mode, the avalanche is allowed to develop to a level that saturates the device, resulting in high detection efficiencies but a current that is independent of the number of photons initiating the avalanche. Recently, a practical scheme for PND has been demonstrated for InGaAs APDs operating at telecom wavelengths.\cite{kardynal08}  This is implemented by gating the device with sub-nanosecond voltage pulses,\cite{yuan07} which are sufficiently short so as for photo-induced avalanches \textit{not} to saturate the device. However, these APDs are insensitive to photons at wavelengths that are currently most commonly used for wider applications in applied physics.

To bridge this gap, we demonstrate here an efficient PND using a fast-gated Si-APD. Operating close to room temperature we obtain a detection efficiency of 73.8\%, giving an avalanche detection probability of 91.1\% for absorbed photons. The device is able to resolve avalanches due to 0, 1, 2, 3 and 4 photons. This performance makes the detection scheme highly attractive for \emph{single-shot} PND that is urgently required for quantum information processing.

\begin{figure}[b]
\centering\includegraphics[width=.85\columnwidth]{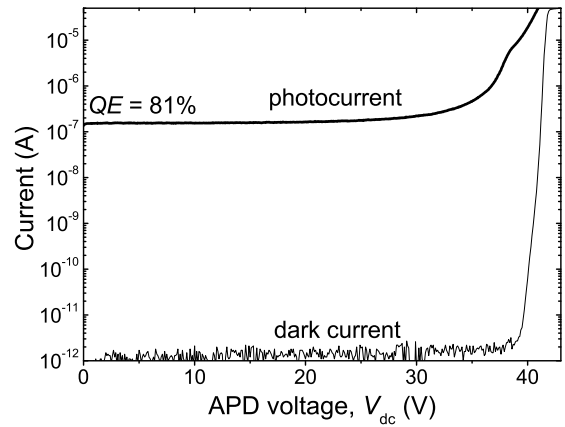}
\caption{Photo- and dark currents vs dc bias. For the photocurrent measurement, the device is illuminated with 0.4~$\mu$W at 600~nm. }
\label{fig:iv}
\end{figure}

For this study we used a Si-APD that was pigtailed with a multi-mode fiber for optical coupling. Cooled thermo-electrically to the operation temperature of --30~$^\circ$C, its breakdown voltage is measured to be 41.7~V at a dark current of 10~$\mu$A (Fig.~\ref{fig:iv}). By measuring the photosensitivity of the APD at dc biases far below the breakdown voltage, for which the gain is unity (Fig.~\ref{fig:iv}), the external $QE$ of the complete system was determined to be $81\pm1$\% at a wavelength of 600~nm. This represents the highest obtainable single photon detection efficiency ($\eta$) with this device at this wavelength. The quantum efficiency can be improved, to a value close to unity, by optimizing the optical coupling, reducing reflection at the front surface and increasing the thickness of the absorbing layer in the APD.  The surface microstructure can also be engineered to enhance the absorption.\cite{wu01}

\begin{figure}[t]
\centering\includegraphics[width=.88\columnwidth]{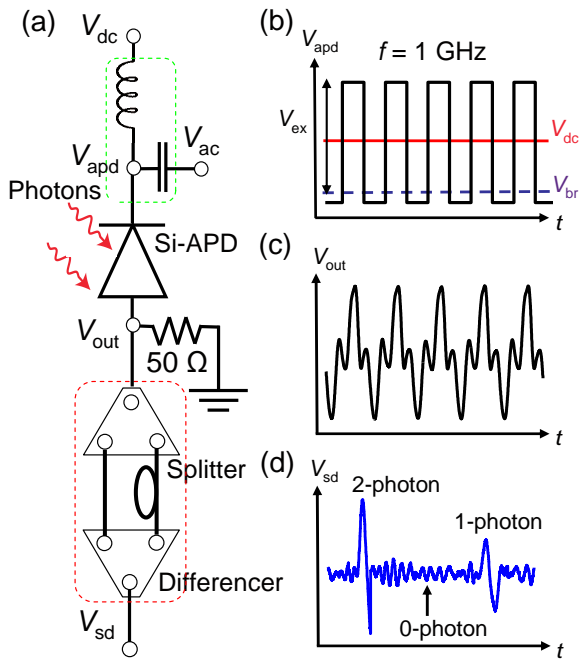}
\caption{(a) Schematic of the Si-APD biasing and measurement electronics;  (b) Gating voltage, $V_{apd}$, supplied to the Si-APD; (c) Output voltage signal from the Si-APD; (d) Self-differencer voltage output, $V_{sd}$. }
\label{fig:sd}
\end{figure}

A schematic of the set-up used to characterize the Si-APD for single photon detection is shown in Fig.~\ref{fig:sd}(a).  A bias voltage consisting of a fixed dc component, $V_{dc}$, and a square wave with a frequency of 1~GHz, $V_{ac}$, [Fig.~\ref{fig:sd}(b)] is applied to the APD so as to periodically produce an excess bias above the breakdown voltage ($V_{br}$), stimulating the detector into a single-photon sensitive state. $V_{ac}\sim10$~V was used throughout the experiment. Photon-induced avalanches cause a current through the APD that is sensed as a voltage across the 50~$\Omega$ resistor.  The avalanches are subsequently quenched during the low voltage part of the detector bias cycle, restricting the duration of the avalanche to be less than 500~ps.

The alternating bias induces a strong background signal [Fig.~\ref{fig:sd}(c)] across the 50~$\Omega$ resistor due to the rapid charging and discharging of the APD.  In order to detect photon-induced avalanches, this capacitive component is removed off the APD response with a self-differencing circuit,\cite{yuan07} which subtracts the gated APD output from that in the preceding period.  This has the effect of cancelling the periodic background, leaving only the response to avalanches.  The output signal of the self-differencer is amplified by a 20-dB broadband amplifier [not shown in Fig.~\ref{fig:sd}(a)] before measurements by an oscilloscope or a time-correlated photon counting system. Figure~\ref{fig:sd}(d) presents a typical output from the device captured on an oscilloscope.  We show later that the output pulses of different amplitude can be attributed to avalanches stimulated by different numbers of incident photons.

\begin{figure}[b]
\centering\includegraphics[width=.96\columnwidth]{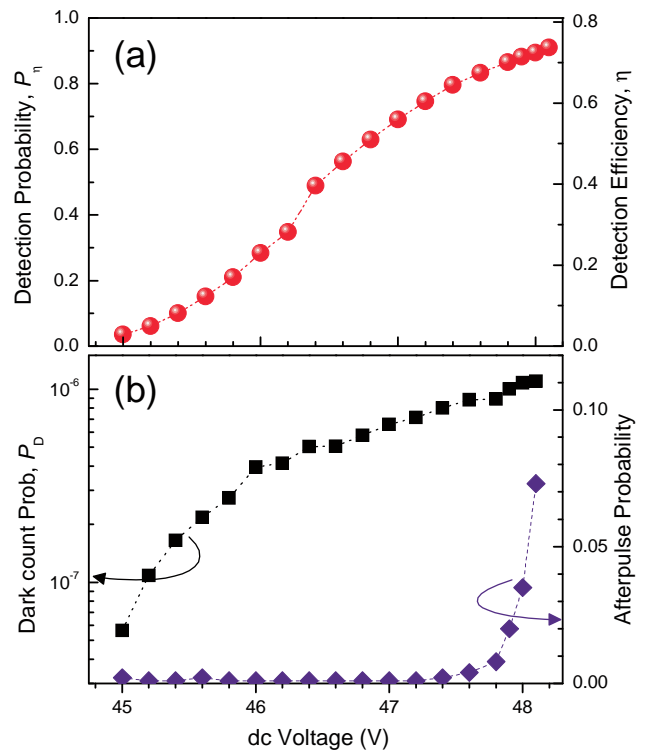}
\caption{(a) Avalanche detection probability ($P_\eta$) for photo-electrons and single photon detection efficiency ($\eta$) as a function of the applied dc bias, $V_{dc}$. Inset shows the dc photosensitivity measured under dark (grey line) and illuminated (black line) conditions; (b) Dark count ($P_D$) and afterpulse ($P_A$) probabilities plotted as a function of the APD dc bias.}
\label{fig:performance}
\end{figure}

The single photon detection efficiency ($\eta$) was first measured as a function of the applied dc bias using the time-resolved single photon counting technique.\cite{yuan07} The optical excitation pulses at 600~nm, obtained using sum frequency generation between a 1550~nm pulsed laser and a 980~nm continuous-wave pump, were synchronized at 1/64 of the gate frequency (1~GHz/64) and attenuated to an average flux of $\mu$=0.033 photons/pulse using a calibrated optical power meter. In determining $\eta$, the dark and afterpulse counts were carefully removed.  As shown in Fig.~\ref{fig:performance}(a), $\eta$ increases with the applied bias to a maximum value of 73.8\%. Using the measured $QE = 81\%$, we are able to obtain an avalanche detection probability ($P_{\eta}=\eta / {QE}$) of 91.1\%, \textit{i.e.} the probability that an absorbed photon produces a detectable avalanche. This suggests that very high detection efficiencies ($\eta>$90\%) may be achieved using our detection scheme with an APD when the external quantum efficiency is optimized.

\begin{figure}[b]
\centering\includegraphics[width=.95\columnwidth]{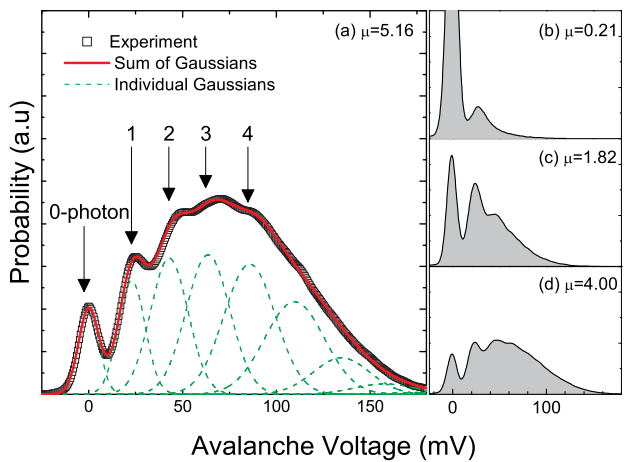}
\caption{(a) Probability distribution (symbols) for avalanche amplitudes measured for a incident photon flux of $\mu = 5.16$. The solid line is a fit using the sum of individual Gaussians (dashed lines) to represent photon number states. (b),(c) and (d) Probability distributions for avalanche amplitudes measured for different photon fluxes of $\mu$ = 0.21, 1.82, and 4.00, respectively.}
\label{fig:pnr}
\end{figure}

As usual, a trade-off is present between the detection efficiency and error rate.  Figure~\ref{fig:performance}(b) shows that the probability of dark counts, \textit{i.e.}, false signals occurring spontaneously in the absence of illumination, increases with the dc bias.  However, the measured dark count probability remains relatively low ($<1.1 \times 10^{-6}$/gate) even for the highest detection efficiencies. APDs also exhibit `afterpulses' [also shown in Fig.~\ref{fig:performance}(b)], which are spurious counts arising from avalanches triggered by re-emission of trapped avalanche charges. Afterpulsing is negligible for a dc voltage of 47.6~V or below. The afterpulse probability increases sharply when further raising the dc bias. This rapid increase in the afterpulse rate occurs because the APD quenching voltage approaches the breakdown condition. This gives rise to a large number of residual unquenched avalanche carriers, which can stimulate secondary additional avalanches.  By using a square wave with higher amplitude than was possible here, we expect to be able to reduce significantly the afterpulse probability, even for avalanche detection probabilities ($P_\eta$) approaching unity. For the present gating conditions, an afterpulse probability of 7.5\% for a corresponding avalanche detection probability of 91.1\% is already very attractive for applications.

To evaluate the PND capability, a fast oscilloscope is used to sample the distribution of output voltage pulses.  Figure~\ref{fig:pnr}(a) shows the measured probability distribution (symbols), acquired from around $5 \times 10^6$ samples for an incident photon flux $\mu = 5.16$, revealing that the APD produces signals of distinctly different amplitudes. These are ascribed to avalanches stimulated by different numbers of photons. The peak at 0~mV reflects the background noise from gates in which no photon was detected. The feature around 22.4~mV is due to the absorption of single photons and those at 43.1~mV, 63.8~mV and 85.5~mV correspond to $N=$ 2, 3 and 4 photons respectively. A good fit (solid line) to the measured data is obtained by modeling each photon number state with a Gaussian (dashed line), assuming the widths of which were scaled by $N^{0.5}$ to reflect the statistical broadening due to avalanche noise. The occurrence probabilities for each photon number is found to be in good agreement with the Poissonian statistics of the light source for a detected flux of 3.7.

The assignment of detected photon number is further confirmed by the illumination flux dependence.  Figures \ref{fig:pnr}(b--d) shows the evolution of the avalanche voltage distribution, measured for different incident photon fluxes. As expected, for $\mu$ = 0.21, signals due to the detection of single photons forms a distinctive peak beside the dominant 0-photon `noise' peak. Signals from 2 or more photons are not clearly visible. As the photon flux increases, the contribution due to 2- and higher photon events increases. The evolution of the area of the photon number peaks is found to be consistent with Poissonian statistics for all incident fluxes.

\begin{figure}[t]
\centering\includegraphics[width=.7\columnwidth]{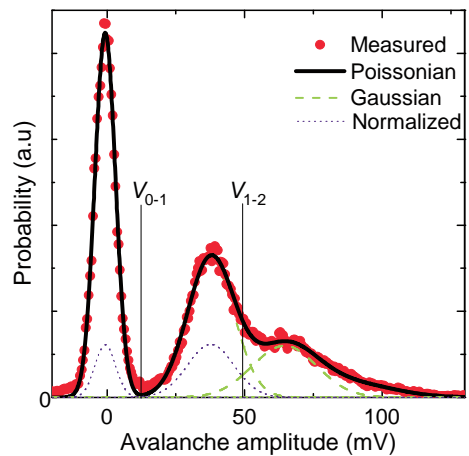}
\caption{ Avalanche probability distributions for a detected flux of 0.92. The dashed curves depict number states corresponding to \emph{N} = 0 - 2, assuming Poissonian statistics of the source, were normalized to reflect a uniform photon distribution (dotted curves). Discrimination levels (thin solid lines) were placed at the crossing between adjacent peaks.}
\label{fig:error}
\end{figure}

We now consider the PND capability of the device for the Bell-state measurements\cite{zukowski93} used in quantum information processing, which require that we distinguish between photon number states with \emph{N} = 0, 1 and $\geq2$.  Figure~\ref{fig:error} shows the output pulse distribution obtained for a moderate detected flux of 0.92, giving rise to significant contributions from only 0, 1 and 2-photon events.  The error in determining the photon number from the measured avalanche voltage, $\varepsilon_N$, was calculated from the numerical overlap between the Gaussians used to fit the photon number states (dashed curves), normalized to reflect a uniform photon number distribution (dotted curves). To retain the detection efficiency we placed single discrimination levels at the crossing between successive photon numbers, for which we obtain photon number detection errors of $\varepsilon_0$ = 0.2 \%, $\varepsilon_1$ = 12.2 \% and $\varepsilon_2$ = 6.95 \%.  

Finally we comment upon the origin of the PND in Si-APDs in which the avalanches are quenched before saturation. The excess noise of the avalanches, quantified as $\langle V^2\rangle / \langle V \rangle ^2$ with $V$ as the avalanche amplitude, is found to be $1.175$ for the 1-photon peak in Fig.~\ref{fig:error}. This value is smaller than 2, which is the theoretical minimum expected for macroscopic avalanche multiplication.\cite{kim99} Such a reduction in the excess noise could arise from two mechanisms: local saturation \textsf{or} intrinsic low-noise avalanche multiplication. Further experiments are required to clarify this issue although it has recently been shown that InGaAs APDs can resolve the photon number for tightly-focused optical excitation.\cite{yuan10}

In conclusion, Si-APDs under short excess bias pulses can be used for photon number detection with high detection efficiency in a practical, thermo-electrically cooled device. The detector is attractive for quantum information applications requiring single-shot photon number detection at high repetition rates.

The authors thank Hamamatsu Photonics for providing some of the devices used in this work, and the EU for partial funding under the FP7 Integrated Project QEssence.



\end{document}